\documentclass{article}

\usepackage[letterpaper,margin=1in]{geometry}
\usepackage{amsmath}
\usepackage{amsfonts}
\usepackage{graphicx}

\newcommand\leqn[1]{\label{eqn:#1}}
\newcommand\eqn[1]{(\ref{eqn:#1})}

    \def\QED{\hfill $\Box$ \medskip}

\newcommand{\NN}{{\mathbb N}}
\newcommand{\PP}{{\mathbb P}}
\newcommand{\RR}{{\mathbb R}}
\newcommand\ind{{\bf 1}}
\newcommand{\bP}{{\bf P}}
\newcommand\kld[2]{D_{\mathrm{KL}}\left({#1}\|{#2}\right)}
\newcommand\given{\,|\,}
\newcommand\abs[1]{\left\vert{#1}\right\vert}
\newcommand\norm[1]{\left\Vert{#1}\right\Vert}
\newcommand\corr{\mathop{\rm corr}}

\title{Markovianness and Conditional Independence in Annotated Bacterial DNA}
\author{Andrew Hart% 
\thanks{Centro de Modelamiento Matem\'atico, Universidad de Chile.}
\and
Servet Mart{\'\i}nez%
\thanks{Centro de Modelamiento Matem\'atico and 
Departamento de Ingenier\'ia Matem\'atica, Universidad de Chile.}
}
\date{\today}

\def\leqn#1{\label{eqn:#1}}
\def\eqn#1{(\ref{eqn:#1})}

\begin{document}

\maketitle

\begin{abstract}
We explore the probabilistic structure of DNA in a number of bacterial genomes
and conclude that a form of Markovianness is present at the boundaries between
coding and non-coding regions, that is, the sequence of START and STOP codons
annotated for the bacterial genome.  This sequence is shown to satisfy a
conditional independence property which allows its governing Markov chain to be
uniquely identified from the abundances of START and STOP codons.  Furthermore,
the annotated sequence is shown to comply with Chargaff's second parity rule at the codon
level.
\end{abstract}

\noindent {\footnotesize

%\bigskip
%\noindent 2010 MSC:  62G10; 62M07; 62P10; 92D20.
}

\bigskip

\section{Introduction}

The strands of DNA composing the genome of an organism are 
segmented along their lengths into two different types of region. 
The first of these are genic regions or genes, whose contents 
can be transcribed into messenger RNA which is
in turn translated into aminoacid polymers for further folding 
and combining to form proteins. In contrast, the remaining 
intergenic regions contain information necessary for 
activities such as the regulation of gene expression and the
management of metabolic networks and controlling cellular processes.

\smallskip

In this article, we consider the boundaries of these regions and the structure
they manifest in the genomes of prokaryotes, principally bacteria.  More
precisely, we seek to uncover the presence of Markovian phenomena at the
interface between genic and intergenic regions.  It has been observed by a
number of authors~\cite{bouaynaya&schonfeld2008,li&kaneko1992,peng&etal1992}
that non-coding regions of chromosomal DNA sequences exhibit long-range
dependence in correlation with respect to the distance between loci on the
strand while coding regions demonstrate short-range dependence.  On the other
hand, \cite{Fedorov&etal2002} reported a power-law decrease in the correlation
between codons in coding regions, which precludes the localized dependence
structure characteristic of Markovianness.

In contrast, we have observed Markovian behaviour at the boundaries between
coding and non-coding regions.  These boundaries are marked by START and STOP
codons.  In the next section, we define what it means for a sequence to be
Markovian and describe a general test for Markovianness introduced
in~\cite{hart&martinez2011}.  We apply this test to the sequences of START and
STOP codons derived from 13 bacterial DNA sequences and conclude that the
annootated STARTs and STOPs constitute a Markov chain.  In addition, we present
less rigorous evidence based on two measures of deviation from Markovianness
which strongly supports the hypothesis that the sequence of STARTs and STOPs is
indeed Markovian.

In Section~\ref{sec:further.struct}, we examine the structure of the START/STOP
Markov chain more deeply and conclude with the aid of entropy and the Kullback-
Leibler divergence that the sequence of START and STOP codons annotated for the
13 chosen bacterial DNA sequences are conditionally independent, a property
which imposes a very precise and simple probabilistic structure on the region
boundaries.  

Finally, we conclude with the observation that each kind of annotated START and
STOP codon appears on the primary and complementary strands with the same
frequency.  this means that the annotated START and STOP codons of a genomic
sequence essentially satisfy Chargaff's second parity rule.  This is notable for
a number of reasons.  Firstly, Chargaff's second parity rule is a symmetry
condition that is generally associated with nucleotide sequences rather than codon
sequences.  Secondly, it is the first time of which we are aware that Chargaff's
second parity rule has been observed in annotated data and lastly the quantity
of annotated data is at the lower limit of the amount generally considered statistically necessary to find
compliance with Chargaff's second parity rule.

\section{Markovianness in strand structure}

A DNA strand essentially comprises a sequence of regions which alternate
between genic zones, which are made up  of a coding sequence initiated by a
START codon, and intergenic zones, which contain no codon instructions for
manufacturing proteins.  A gene (genic zone) is generally considered 
to be a sequence of codons (trinucleotides) which begins with a START 
codon and which ends with one of the three immutable STOP codons 
TAA, TAG or TGA.  Since we are only considering prokaryotes here, 
we do not have to contend with the presence
of introns within genic regions.  Typical START codons for bacteria 
include ATG, GTG and TTG, but their may be others as well depending 
on the organism.  As codons comprise three nucleic acid bases, each 
genic region may appear in any of 3 possible reading frames.  Although 
START codons can vary between organisms,
the set of START codons never overlaps the set of STOP codons.  For our
purposes, we shall view a region as being any sequence of bases 
that begins with a START codon or STOP codon.  
Regions commencing with a START codon will be
genic while those beginning with a STOP codon will be intergenic.

\medskip

As noted above, Markovian processes are not the most appropriate vehicle
for modelling sequences of DNA, 
despite the extensive and successful 
use of Markovian concepts in gene identification and anotation.  Markovianness is a property 
of a system which captures the idea that when a change of state occurs, 
the new state only depends on the system's state immediately prior to 
the change and not on any other antecedent states.  In a time series, 
Markovianness means that the future and the past are independent of 
each other given the present state of the series. In a DNA sequence, 
Markovianness can be interpreted as saying that given knowledge of a 
base at a particular position in the sequence, the nucleotides that precede 
the position are independent of those that follow it. For many modelling 
problems, an assumption of Markovianness is perfectly reasonable, even 
if it is not in fact true.  In such cases, Markovianness often captures 
enough of the structure of the system to provide a satisfactory
approximation.  However, the complexity of biological systems generlly 
precludes the imposition of such a strong assumption as Markovianness on 
its probabilistic structure.

\medskip

Despite this, we have observed the presence of a restricted form of
Markovianness at the boundaries of regions as we have defined them here.  
We shall present evidence for this Markovianness in two different ways.  
Our chief tool for detecting Markovianness is the test for Markovianness 
for sequences over finite alphabets developed in~\cite{hart&martinez2011}. 
We give a very brief resumé of the test below, before summarizing the 
results of applying it to the 13 sequences.

\subsection{Testing for Markovianness}

A finite Markov chain is a dynamical system which evolves on a finite state
space, say, $I$.  For this brief explanation, we shall think of the chain as
evolving in time.  Thus, the Markov chain produces a sequence of states
$i_0,i_1,\ldots,i_t,\ldots$.  Now, according to the Markov property, state
$i_{t+1}$ only depends on $i_t$ and not on any of the states prior 
to time~$t$.  Thus, $i_{t+1}$ may be viewed as a function of $i_t$ together 
with an external influence variously called the noise, innovation 
or disturbance at time~$t$:
\begin{equation}
\leqn{markov.chain}
i_{t+1} = f(i_t, U_t), \qquad t=0,1,\ldots.
\end{equation}
Here, $U_t$ is the unobservable noise at time~$t$.  The sequence
$U_0,U_1,U_2,\ldots$ must be a sequence of independent and identically
distributed random variables.  
If $u_t$ were to depend on $u_{t-1}$, this would
constitute a violation of the Markov property since $i_{t+1}$ 
would then depend (albeit indirectly) on $U_{t-1}$, as would $i_t$ 
since $i_t=f(i_{t- 1},U_{t-1})$.  We need all the $U_i$'s 
to be identically distributed in order to uncouple the 
mechanism governing the transition from state $i_t$ to $i_{t+1}$
from the particular time~$t$ at which the transition occurs.

The test for Markovianness is based on the fact that for any Markov chain,
the function~$f$ can always be chosen so that the noise sequence 
can be taken to be uniformly distributed on the interval $[0,1]$. 
Suppose that we have a sequence $i_0,i_1,\ldots,i_n$. Then, due to 
\eqn{markov.chain}, there is a limited range of values of $U_t$ that can 
result in state $i_{t+1}$ being observed following state $i_t$. Denote 
this set of values by $F_{-1}(i_{t+1},i_t)\subseteq[0,1]$.
Note that this does not depend on~$t$.  Furthermore, given $i_t$ 
and $i_{t+1}$, $U_t$ is uniformly distributed over the set 
$F_{-1}(i_{t+1},i_t)$. Consequently, the conditional distribution of $U_t$ 
given $i_t$ and $i_{t+1}$ is known and surrogates $U_0',\ldots,U_n'$ for 
the sequence $U_0,\ldots,U_n$ can be obtained by simulating values from 
the conditional distributions.  Then, if the sequence $U_0',\ldots,U_n'$ 
is independently and identicaly distributed uniformly on $[0,1]$, it 
is consistent with $i_0,\ldots,i_n$ having been generated by a Markov 
chain.  Consequently, we can exchange the problem of testing the
Markovianness of a sequence for that of testing the independence and 
uniformity of the sequence $U_0',\ldots,U_n'$ and there exist standard 
statistical tests for this.

\medskip

By default we use a collection of tests, with their $p$-values appropriately
adjusted to compensate for multiple testing, to decide whether or 
not a given sequence could have been produced by a finite state Markov 
chain.  Here, we shall use the 
Ljung-Box $q$ test~\cite{ljung&box1976} with 20 lags to test for independence and the 
one-sample Kolmogorov-Smirnov test to test for uniformity of
$U_0',\ldots,U_n'$ \cite[Chapter 9]{degroot1991}.  We used the Holm-Bonferroni method to 
adjust the $p$-values to correct for multiple testing and we accept 
the null hypothesis of Markovianness at a significance level~$\alpha$
if the two adjusted $p$- values are greater than~$\alpha$.

\medskip

We considered the genomes of 11 bacteria which include a total of 
$13$ chromosomal sequences and used coding sequences (cds) annotated 
in GenBank to identify genic and intergenic regions.  In particular, 
we noted the START and STOP codons, as well as the strand on which each 
appears.  The first thing we did was to 
consider the START and STOP codons themselves as a sequence.
For example, the first 8 START/STOP codons appearing on the primary 
strand of escherichia coli K-12 substr.\ MG1655 according to the 
annotation available in GenBank are:  ATG, TGA, ATG, TGA, ATG, 
TAA, ATG, TAA.  We applied the statistical test for Markovianness 
described in the preceding section to the primary strands of a 
small collection of genomes. The results obtained are displayed in 
Table~\ref{tab:1}.  Similarly, Table~\ref{tab:2} shows the results
of applying the same test to the complementary strands of the same 
genomes.

\medskip

The adjusted $p$-values displayed in both tables suggest that the sequence 
of START/STOP codons is Markovian in nature for the genomes tested.

\begin{table}[tbp]
\caption{$p$-values for the Markov test applied to the sequence of START 
and STOP codons on the \textit{primary} strand of 13 bacterial DNA chromosomes. 
$p$-values for the Markov test based on the Ljung-Box $Q$ test for correlation and the Kolmogorov-Smirnov test for uniformity on $[0,1]$ are shown. Numbers in parentheses 
represent the $p$-values adjusted for multiple 
testing of the same genomic sequence using the Holm- Bonferroni method. 
No correction has been applied to account for
the testing of multiple sequences.}
\label{tab:1}
\centering
\begin{tabular}{p{8cm}rrrr}
  \hline
 Chromosome & \multicolumn2c{Ljung-Box Test} & \multicolumn2c{K-S Test} \\
  & $p$-value & Adjusted & $p$-value & Adjusted \\
  \hline
Escherichia coli str. K-12 substr. MG1655 & 0.91 & (0.91) & 0.30 & (0.60) \\
  Helicobacter pylori 26695 chromosome & 0.09 & (0.19) & 0.85 & (0.85) \\
  Staphylococcus aureus subsp. aureus MRSA252 chromosome & 0.36 & (0.72) & 0.95 & (0.95) \\
  Leptospira interrogans serovar Lai str. 56601 chromosome I & 0.23 & (0.47) & 1.00 & (1.00) \\
  Leptospira interrogans serovar Lai str. 56601 chromosome II & 0.86 & (1.00) & 0.69 & (1.00) \\
  Streptococcus pneumoniae ATCC 700669, complete genome. & 0.12 & (0.24) & 0.48 & (0.48) \\
  Bacillus subtilis subsp. spizizenii str. W23 chromosome & 0.82 & (0.82) & 0.08 & (0.17) \\
  Vibrio cholerae O1 str. 2010EL-1786 chromosome 1 & 0.81 & )1.00) & 0.84 & (1.00) \\
  Vibrio cholerae O1 str. 2010EL-1786 chromosome 2 & 0.39 & (0.79) & 0.63 & (0.79) \\
  Propionibacterium acnes TypeIA2 P.acn33 chromosome & 0.10 & (0.20) & 0.86 & (0.86) \\
  Salmonella enterica subsp. enterica serovar Typhi str. P-stx-12 & 0.03 & (0.06) & 0.56 & (0.56) \\
  Yersinia pestis D182038 chromosome & 0.51 & (0.57) & 0.29 & (0.57) \\
  Mycobacterium tuberculosis 7199-99 & 0.53 & (1.00) & 0.54 & (1.00) \\
   \hline
\end{tabular}
\end{table}

\begin{table}[tbp]
\caption{$p$-values for the Markov test applied to the sequence of START 
and STOP codons on the \textit{complementary} strand of 13 bacterial DNA chromosomes.  
$p$-values for the Markov test based on the Ljung-Box $Q$ test for correlation and the Kolmogorov-Smirnov test for uniformity on $[0,1]$ are shown. Numbers in parentheses 
represent the $p$-values adjusted for 
multiple testing of the same genomic sequence using the Holm-
Bonferroni method. No correction has been applied to 
account for the testing of multiple sequences.}
\label{tab:2}
\bigskip
\centering
\begin{tabular}{p{8cm}rrrr}
  \hline
 Chromosome & \multicolumn2c{Ljung-Box Test} & \multicolumn2c{K-S Test} \\
  & $p$-value & Adjusted & $p$-value & Adjusted \\
 \hline
Escherichia coli str. K-12 substr. MG1655 & 0.61 & (1.00) & 0.70 & (1.00) \\ 
  Helicobacter pylori 26695 chromosome & 0.57 & (0.57( & 0.25 & (0.50) \\ 
  Staphylococcus aureus subsp. aureus MRSA252 chromosome & 0.49 & (0.98) & 0.59 & (0.98) \\ 
  Leptospira interrogans serovar Lai str. 56601 chromosome I & 0.11 & (0.22) & 0.42 & (0.42) \\ 
  Leptospira interrogans serovar Lai str. 56601 chromosome II & 0.50 & (0.50) & 0.14 & (0.27) \\ 
  Streptococcus pneumoniae ATCC 700669, complete genome. & 0.59 & (1.00) & 0.82 & (1.00) \\ 
  Bacillus subtilis subsp. spizizenii str. W23 chromosome & 0.89 & (1.00) & 0.75 & (1.00) \\ 
  Vibrio cholerae O1 str. 2010EL-1786 chromosome 1 & 0.60 & (1.00) & 0.97 & (1.00) \\ 
  Vibrio cholerae O1 str. 2010EL-1786 chromosome 2 & 0.81 & (1.00) & 0.92 & (1.00) \\ 
  Propionibacterium acnes TypeIA2 P.acn33 chromosome & 0.93 & (1.00) & 0.51 & (1.00) \\ 
  Salmonella enterica subsp. enterica serovar Typhi str. P-stx-12 & 0.14 & (0.29) & 0.96 & (0.96) \\ 
  Yersinia pestis D182038 chromosome & 0.10 & (0.19) & 0.81 & (0.81) \\ 
  Mycobacterium tuberculosis 7199-99 & 0.24 & (0.49) & 0.92 & (0.92) \\ 
   \hline
\end{tabular}
\end{table}

\subsection{Measuring deviation from Markovianness}

We can present further evidence to support the hypothesis of Markovianness
of the sequences of START and STOP codons.  Though less rigourous than a
statistical hypothesis test, we have found a statistic which is sensitive to
deviations from Markovianness in sequences of finite symbols.  We shall first
describe this measure and demonstrate it using simulated Markovian and non-
Markovian data.  Then, we shall compare the measure for annotated START/STOP
codons in bacterial DNA sequences  with the same measure applied to simulations
of Markovian and non-Markovian sequences possessing similar statistical
properties to those derived from the annotation data.

\medskip

Let $(X_t: t\in\NN)$ be a sequence of symbols in~$I$. 
Here, $I$ is the set of START/STOP codons of a bacterial genome.  

\medskip

If $(X_t)$ has the Markov property, this  means that
\begin{equation}
\leqn{markov.prop}
\PP(X_{t+1}=k \given X_t=j, X_{t-1}=i, X_{t-2}=i_{t-2},
\ldots, X_0=i_0) = \PP(X_{t+1}=k\given X_t=j),
\end{equation}
for all integers $t>0$.  By multiplying both sides of \eqn{markov.prop} by
$\PP(X_t=j,X_{t-1}=i, X_{t-2}=i_{t-2},\ldots,X_0=i_0)$ and summing over
$i_0,i_1,\ldots,i_{t-2}\in I$, it can be seen that the Markov property implies
\begin{align*}
& \PP(X_{t-1}=i, X_t=j,X_{t+1}=k) \\
=& \PP(X_{t+1}=k \given X_t=j,X_{t-1}=i) 
\PP(X_t=j,X_{t-1}=i) = \PP(X_{t+1}=k \given X_t=j) \PP(X_t=j,X_{t-1}=i) \\
=& \frac{\PP(X_{t-1}=i,X_t=j)\PP(X_t=j,X_{t+1}=k)}{\PP(X_t=j)}.
\end{align*}
Under stationarity, the above does not depend on~$t$, so we can 
write it in the more compact form
$$
\PP([ijk]) = \frac{\PP([ij])\PP([jk])}{\PP([j])},
$$
where $[i]$, $[ij]$ and $[ijk]$ denote the cylinder sets of 
length one, two and three symbols respectively.
Therefore, when $(X_t)$ is a Markovian sequence, $M_3(i,j,k)=0$, 
for all $i,j,k\in I$, where
$$
M_3(i,j,k) 
= \PP([ijk]) - \frac{\PP([ij])\PP([jk])}{\PP([j])}.
$$
It is straightforward to estimate the quantities 
$M_3(i,j,k)$ for a sequence by counting the occurrences of 
single codons, pairs of codons and groups of three
codons.  If $N_i$, $N_{ij}$ and $N_{ijk}$ denote the 
frequencies of $i$, $ij$ and $ijk$ respectively, 
then $M_3(i,j,k)$ can be estimated by
$$
\widehat M_3(i,j,k) = \frac{N_{ijk}}n - \frac{N_{ij}N_{jk}}{nN_j},
$$
where~$n$ is the length of the sequence. For purposes of calculating $N_i$,
$N_{ij}$ and $N_{ijk}$, we treat the sequence $X_t)$ as circular so that
$\sum_{i\in I}N_i = \sum_{i,j\in I}N_{ij} = \sum_{i,j,k\in I}N_{ijk}=n$.  
This also means that $N_{ij}=\sum_{k\in i}N_{ijk}$ and 
$N_i=\sum_{j\in I}N_{ij}$.

\medskip

Now, $\widehat M_3=(\widehat M_3(i,j,k): i,j,k\in I)$ is a collection of $|I|^3$ values,
each of which is the deviation by the corresponding cylinder $[ijk]$ from
Markovianness.  Note that, because sequences of START/STOP codons alternate
between START codons and STOP codons, many elements of $M_3$ and $\widehat M_3$ will
be zero. For example, all but the last bacterial sequence listed in
Tables~\ref{tab:1} and~\ref{tab:2} has three START codons $\{ATG,GTG,TTG\}$ and
3 STOP codons $\{TAA,TAG,TGA\}$.  The last bacteria, Yersinia pestis D182038,
employs an extra START codon, $CTG$.  Thus, $|I|=6$ in general and $\widehat M_3$
will have 216 elements, of which at least 162 will be zero. The mean of $\widehat
M_3$ is
\begin{align*}
\bar{\widehat M}_3
&= \frac1n\sum_{i,j,k\in I}\frac{N_{ijk}}n - \frac1n\sum_{i,j,k\in I}\frac{N_{ij}N_{jk}}{nN_j} \\
&= \frac n{n^2} - \frac1n\sum_{i,j\in I}\frac{N_{ij}N_j}{nN_j} \\
&= \frac1n - \frac1n\sum_{i,j\in I} \frac{N_{ij}}n 
= 1/n - 1/n = 0.
\end{align*}

Through empirical experimentation, we found that the sample 
standard deviation of $\widehat M_3$ provides a statistic that is 
responsive to departures from Markovianness:
\begin{align*}
S_3 &= \sigma(\widehat M_3) = \sqrt{\frac1{n-1}\sum_{i,j,k\in I}
\left(\widehat M_3(i,j,k)-\bar{\widehat M}_3(i,j,k)\right)^2} \\
&= \sqrt{\frac1{n-1}\sum_{i,j,k\in I}\widehat M_3^2(i,j,k)}.
\end{align*}

\begin{figure}[tbp]
\centering
\includegraphics[width=5.8in]{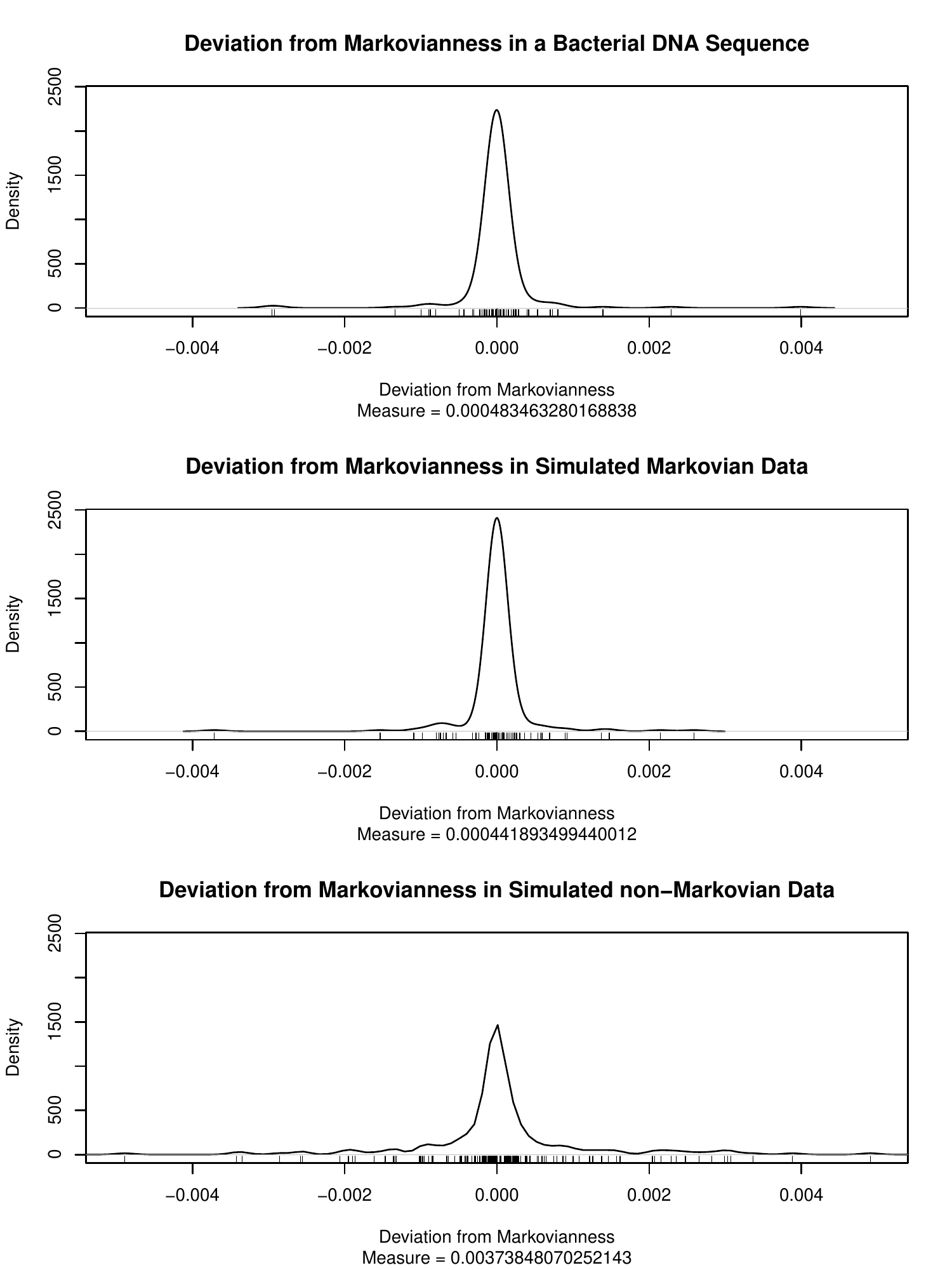}

\caption{Empirical evidence of the efficacy of using the standard 
deviation as a measure of the degree by which a sequence deviates 
from the Markov property. The top figure pertains to the START/STOP 
codons annotated on the primary strand of escherichia coli K-12.  
The deviations are marked on the x-axis while the curve
represents a kernel density estimate of the deviations.  
The middle plot illustrates the same thing using a simulated Markov chain.  
The bottom plot was produced using non-Markovian simulations of 
latent AR$(2)$ processes.}
\label{fig:1}
\end{figure}

Figure~\ref{fig:1} displays a kernel density estimate for $\widehat M_3$ in three
cases. The first case shows the density of $\widehat M_3$ for the sequence of
START/STOP codons annotated on the primary strand of the escherichia coli K-12
genome. Let us denote this sequence by $X^1$. The second case shows the density
estimated for a sequence~$X^2$ of START/STOP codons simulated from a Markov
chain using a transition matrix estimated from $X^1$. The idea is that $X^1$ and
$X^2$ be statistically the same for single codons and pairs of consecutive
codons so that $\widehat M_3$ only highlights the kind of mechanism, Markovian or
non-Markovian, driving the process.  In the third case, a latent AR$(2)$ process
was simulated using the following scheme.

\medskip

Let $(Z_k: k=0,1,\ldots,n)$ be an AR$(2)$ process with 
autoregressive coefficients $\lambda_1$ and $\lambda_2$, that is:
$$
Z_t = \lambda_1Z_{t-1} + \lambda_2Z_{t-2} + \epsilon_t,
$$
where the innovations $\epsilon_t$ are independently and 
identically distributed normal random variables with mean~$0$ 
and variance~$\sigma^2$.  The process~$Z$
is stationary if and only if the parameters satisfy the conditions
\begin{equation}
\leqn{ar2.boundary}
\lambda_2>-1, \qquad \lambda_2+\lambda_1<1 \mbox{ and } 
\lambda_2-\lambda_1<1.
\end{equation}
Note that~$Z$ is a Markov chain if and only if $\lambda_2=0$ and an i.i.d.
process if and only if $\lambda_2=\lambda_1=0$.

\medskip

Next, let $q_Z(p)$ denote the quantile function of~$Z$ , that is,
$$
q_Z(p) = \max\left\{z\in \RR \;:\; \biggl(\frac1n\sum_{t=0}^n\ind_{Z_t\leq z}\biggr) \leq p\right\}.
$$

Finally, we define the stochastic process $(Y_t: t=0,1,\ldots,n)$.  
To do this, we require that the symbols in~$I$ are ordered in some way. 
The order does not matter, we merely need to be able to say for 
$i,j\in I$ that either~$i$ comes before~$j$ or~$j$ comes before~$i$. 
Let $\underline i$ denote the symbol in~$I$ that comes before all 
others in~$I$. Then, the latent AR$(2)$ process is then defined as 
$$
Y_t = \left\{\begin{array}{ll}
\underline i & \mbox{if } Y_t \leq q_Z(\pi_{\underline i}), \\
i, &  \mbox{if } q_Z\left(\sum_{j<i}(\pi_j)\right) <
Z_t \leq q_Z\left(\sum_{j\leq i}\pi_j)\right).
\end{array}\right.
$$
Due to how~$Y$ has been constructed, $\pi$ is its invariant state 
distribution. Also, $Y$ will be Markovian if and only if~$Z$ is 
Markovian (equivalently, $\lambda_2=0$).

\medskip

In the third case, we simulated a sequence~$X^3$ of START and STOP codons from 
the latent AR$(2)$ process described above.  In order to obtain a non-Markovian
sequence with the same distribution of symbols as $X^1$, we set $\lambda_1=-0.2$
and $\lambda_2=0.4$, and estimated~$\pi$ from $X^1$ .

\medskip

In Figure~\ref{fig:1}, the densities of the deviations $\widehat M_3$ for the
sequences derived from escherichia coli and the Markov chain simulation 
are fairly similar. Their statistics $S_3$ are also comparable.  
In contrast, the density of $\widehat M_3$ for the non-Markovian simulation 
has much longer tails and exhibits much greater dispersion.  
The x-axis of the third plot in the figure has been truncated to the 
interval $[0.005, 0.005]$ to maintain clarity and allow for easy 
comparison with the other two densities.  All three graphs have
been plotted on the same scale also for this reason.  Prior to truncation, 
the density of $\widehat M_3$ for the non-Markovian sample spanned the 
interval $[-0.0195, 0.0327]$ and 14 data points are omitted by the 
truncation.  The measure $S_3$ for the simulated latent AR$(2)$ process 
is almost an order of magnitude larger than it is for the other two cases.

\begin{table}[tbp]
\caption{A measure of Markovianness based on a trinucleotide analysis 
applied to the annotated START and STOP codons on the \textit{primary} 
strand sequences of bacterial genomes, together with Markovian 
and non-Markovian  simulations.  The Markovian simulations have statistically equivalent 
dinucleotide distributions to the annotated STARTs/STOPs while the 
non-Markovian simulations have the same
mononucleotide distributions as their bacterial counterparts.}
\label{tab:3}
\bigskip
\centering
\begin{tabular}{p{7cm}rrrr}
  \hline
Chromosome & \multicolumn3c{Measure from} & Number of \\
& Genome & Markovian & Non-Markovian & STARTs \\
& & Simulations & Simulations & and STOPs \\
  \hline
   Escherichia coli str. K-12 substr. MG1655 
 & 0.000483 & 0.000379 & 0.003977 & 4058 \\ 
  Helicobacter pylori 26695 chromosome
& 0.000617 & 0.000798 & 0.003590 & 1528 \\ 
  Staphylococcus aureus subsp. aureus 
\newline MRSA252 chromosome
& 0.000654 & 0.000487 & 0.003731 & 2560 \\ 
  Leptospira interrogans serovar Lai str. 56601 chromosome I
& 0.000695 & 0.000527 & 0.003230 & 3626 \\ 
  Leptospira interrogans serovar Lai str. 56601 chromosome II 
& 0.002731 & 0.001805 & 0.004268 & 320 \\ 
  Streptococcus pneumoniae \newline ATCC 700669 
  & 0.000801 & 0.000579 & 0.003776 & 1910 \\ 
  Bacillus subtilis subsp. spizizenii str. W23 chromosome, complete 
& 0.000373 & 0.000495 & 0.003509 & 3844 \\ 
  Vibrio cholerae O1 str. 2010EL-1786 chromosome 1
& 0.000715 & 0.000543 & 0.003598 & 2750 \\ 
  Vibrio cholerae O1 str. 2010EL-1786 chromosome 2
& 0.000560 & 0.000836 & 0.003738 & 1162 \\ 
  Propionibacterium acnes TypeIA2 \newline P.acn33 chromosome 
& 0.000882 & 0.000513 & 0.003983 & 2234 \\ 
  Salmonella enterica subsp. enterica serovar Typhi str. P-stx-12 
& 0.001453 & 0.000419 & 0.003632 & 4806 \\ 
  Yersinia pestis D182038 chromosome 
& 0.000592 & 0.000511 & 0.003502 & 3430 \\ 
  Mycobacterium tuberculosis 7199-99 
& 0.000483 & 0.000460 & 0.002453 & 4006 \\ 
\hline
\end{tabular}
\end{table}

\begin{table}[tbp]
\caption{A measure of Markovianness based on a trinucleotide analysis 
applied to the annotated START and STOP codons on the \textit{complementary} 
strand sequences of bacterial genomes, together with Markovian 
and non-Markovian  simulations.  The Markovian simulations 
have statistically equivalent dinucleotide distributions to the 
annotated STARTs/STOPs while the non-Markovian simulations have the 
same mononucleotide distributions as their bacterial counterparts.}
\label{tab:4}
  \bigskip
\centering
  \begin{tabular}{p{7cm}rrrr}
  \hline
Chromosome & \multicolumn3c{Measure from} & Number of \\
& Genome & Markovian & Non-Markovian & STARTs \\
& & Simulations & Simulations & and STOPs \\
\hline
   Escherichia coli str. K-12 substr. MG1655 
& 0.000786 & 0.000353 & 0.004120 & 4284 \\ 
  Helicobacter pylori 26695 chromosome 
& 0.001364 & 0.000766 & 0.003741 & 1606 \\ 
  Staphylococcus aureus subsp. aureus 
\newline MRSA252 chromosome
& 0.000748 & 0.000455 & 0.003839 & 2730 \\ 
  Leptospira interrogans serovar Lai str. 56601 chromosome I 
& 0.000736 & 0.000571 & 0.003334 & 3192 \\ 
  Leptospira interrogans serovar Lai str. 56601 chromosome II
& 0.001676 & 0.001921 & 0.004303 & 266 \\ 
  Streptococcus pneumoniae \newline ATCC 700669 
& 0.000587 & 0.000588 & 0.003635 & 2070 \\ 
  Bacillus subtilis subsp. spizizenii str. W23 chromosome
& 0.000521 & 0.000468 & 0.003460 & 4276 \\ 
  Vibrio cholerae O1 str. 2010EL-1786 chromosome 1
& 0.001121 & 0.000526 & 0.003552 & 2860 \\ 
  Vibrio cholerae O1 str. 2010EL-1786 chromosome 2
& 0.000750 & 0.000976 & 0.003687 & 918 \\ 
  Propionibacterium acnes TypeIA2 
\newline P.acn33 chromosome, complete & 0.000949 
& 0.000528 & 0.003908 & 2232 \\ 
  Salmonella enterica subsp. enterica serovar Typhi str. P-stx-12 
& 0.000705 & 0.000417 & 0.003508 & 4574 \\ 
  Yersinia pestis D182038 chromosome 
  & 0.000644 & 0.000497 & 0.003606 & 3810 \\ 
  Mycobacterium tuberculosis 7199-99  
& 0.000509 & 0.000461 & 0.002546 & 3962 \\ 
\hline
\end{tabular}
\end{table}

\bigskip

Table~\ref{tab:3} displays the value of $S_3$ computed on the primary 
strand of 13 bacterial DNA sequences.  The second column shows	 
$S_3$ derived from genome annotation data.  The third and fourth 
columns show the measure of deviation from Markovianness as applied 
to Markovian and non-Markovian sequences respectively simulated 
as described above.  For each of these columns, a sequence of the 
same length as the annotated START/STOP codons was simulated $1000$ times 
and the mean value of $S_3$ over the $1000$ replications is shown
in the table.  For the non-Markovian case, the autoregressive parameters
$\lambda_1$ and $\lambda_2$ were selected uniformly at random 
from the set of values that give rise to a stationary AR$(2)$ process for each simulation.  
It is quite evident that the values of $S_3$ for the annotation data 
and the Markovian simulations are of the same order of magnitude while 
the non-Markovian simulations result in values of $S_3$ that are from 
two times to an order of magnitude greater.  The final column in the table 
shows the length of the sequence of START and STOP codons annotated 
for each of the DNA sequences.  There appears to be no relationship
between the sequence length and any of the measures of deviation from
Markovianness calculated.  Performing the same analysis on the 
complementary strands yields similar results which are shown 
in Table~\ref{tab:4}.

\begin{table}[tbp]
\caption{A measure of Markovianness based on a quadranucleotide 
analysis  applied to the annotated START and STOP codons on the 
\textit{primary} strand sequences of bacterial genomes, together with 
Markovian and non-Markovian  simulations.  The Markovian simulations 
have statistically equivalent dinucleotide distributions to the annotated 
STARTs/STOPs while the non-Markovian simulations have the same
mononucleotide distributions as their bacterial counterparts.}
\label{tab:5}
\bigskip
\c\centering
\begin{tabular}{p{7cm}rrrr}
  \hline
Chromosome & \multicolumn3c{Measure from} & Number of \\
& Genome & Markovian & Non-Markovian & STARTs \\
& & Simulations & Simulations & and STOPs \\
  \hline
Escherichia coli str. K-12 substr. MG1655 
& 0.000199 & 0.000188 & 0.001399 & 4058 \\ 
  Helicobacter pylori 26695 chromosome 
& 0.000408 & 0.000404 & 0.001213 & 1528 \\ 
  Staphylococcus aureus subsp. aureus 
\newline MRSA252 chromosome
& 0.000339 & 0.000253 & 0.001236 & 2560 \\ 
  Leptospira interrogans serovar Lai str. 56601 chromosome I 
& 0.000338 & 0.000266 & 0.001084 & 3626 \\ 
  Leptospira interrogans serovar Lai str. 56601 chromosome II 
& 0.001170 & 0.000904 & 0.001566 & 320 \\ 
  Streptococcus pneumoniae \newline ATCC 700669 
& 0.000380 & 0.000280 & 0.001268 & 1910 \\ 
  Bacillus subtilis subsp. spizizenii str. W23 chromosome
& 0.000249 & 0.000251 & 0.001054 & 3844 \\ 
  Vibrio cholerae O1 str. 2010EL-1786 chromosome 1
& 0.000322 & 0.000270 & 0.001208 & 2750 \\ 
  Vibrio cholerae O1 str. 2010EL-1786 chromosome 2
& 0.000397 & 0.000425 & 0.001228 & 1162 \\ 
  Propionibacterium acnes TypeIA2 P\newline .acn33 chromosome
& 0.000325 & 0.000271 & 0.001427 & 2234 \\ 
  Salmonella enterica subsp. enterica serovar Typhi str. P-stx-12 
& 0.000513 & 0.000206 & 0.001218 & 4806 \\ 
  Yersinia pestis D182038 chromosome 
& 0.000297 & 0.000257 & 0.001101 & 3430 \\ 
  Mycobacterium tuberculosis 7199-99 
& 0.000240 & 0.000212 & 0.000706 & 4006 \\ 
   \hline
\end{tabular}
\end{table}

\begin{table}[tbp]
\caption{A measure of Markovianness based on a quadranucleotide
analysis  applied to the annotated START and STOP codons on the 
\textit{complementary} strand sequences of bacterial genomes, together 
with Markovian and non-Markovian  simulations.  The Markovian simulations 
have statistically equivalent dinucleotide distributions to the annotated 
STARTs/STOPs while the non-Markovian simulations have the same
mononucleotide distributions as their bacterial counterparts.}
\label{tab:6}
\bigskip
\centering
\begin{tabular}{p{7cm}rrrr}
  \hline
Chromosome & \multicolumn3c{Measure from} & Number of \\
& Genome & Markovian & Non-Markovian & STARTs \\
& & Simulations & Simulations & and STOPs \\
  \hline
 	Escherichia coli str. K-12 substr. MG1655 
& 0.000307 & 0.000179 & 0.001372 & 4284 \\ 
  Helicobacter pylori 26695 chromosome 
& 0.000500 & 0.000378 & 0.001157 & 1606 \\ 
  Staphylococcus aureus subsp. aureus 
\newline MRSA252 chromosome
& 0.000259 & 0.000243 & 0.001195 & 2730 \\ 
  Leptospira interrogans serovar Lai str. 56601 chromosome I 
& 0.000369 & 0.000287 & 0.001067 & 3192 \\ 
  Leptospira interrogans serovar Lai str. 56601 chromosome II 
& 0.000857 & 0.000973 & 0.001650 & 266 \\ 
  Streptococcus pneumoniae 
\newline ATCC 700669 
& 0.000302 & 0.000282 & 0.001251 & 2070 \\ 
  Bacillus subtilis subsp. spizizenii str. W23 chromosome
& 0.000244 & 0.000234 & 0.001092 & 4276 \\ 
  Vibrio cholerae O1 str. 2010EL-1786 chromosome 1
& 0.000478 & 0.000266 & 0.001160 & 2860 \\ 
  Vibrio cholerae O1 str. 2010EL-1786 chromosome 2
& 0.000443 & 0.000485 & 0.001274 & 918 \\ 
  Propionibacterium acnes TypeIA2 
\newline P.acn33 chromosome
& 0.000398 & 0.000274 & 0.001367 & 2232 \\ 
  Salmonella enterica subsp. enterica serovar Typhi str. P-stx-12 
& 0.000395 & 0.000216 & 0.001165 & 4574 \\ 
  Yersinia pestis D182038 chromosome 
& 0.000274 & 0.000244 & 0.001182 & 3810 \\ 
  Mycobacterium tuberculosis 7199-99 
& 0.000220 & 0.000213 & 0.000731 & 3962 \\ 
   \hline
\end{tabular}
\end{table}

\medskip

We can also consider Markovianness in terms of quadranucleotides.  In this case, $[ijkl]$ is the cylinder set for quadranucleotide $ijkl$. We define
$$
M_4(i,j,k,l) 
= \PP([ijkl]) - \frac{\PP([ijk])\PP([kl])}{\PP([k])},
\qquad i,j,k,l\in I.
$$
Now, if $X_t$ is be Markovian then $M_4(i,j,k,l)=0$ for all
 $i,j,k,l\in I$.  Once again, we note that for most of the bacteria we examined,
the alternating nature of their sequences of START and STOP codons means that a
minimum of 1134 elements of $M_4=(M_4(i,j,k): i,j,k,l\in I)$ will be zero,
regardless of whether or not the sequence of STARTs and STOPs is Markovian.  In
a manner similar to the case for $M_3$, we can estimate $M_4(i,j,k,l)$ by
$$
\widehat M_4(i,j,k,l) = \frac{N_{ijkl}}n - \frac{N_{ijk}N_{kl}}{nN_k}.
$$
Furthermore, the mean $\bar{\widehat M}_4=0$ and
$$
S_4=\sigma(\widehat M_4) = \sum_{i,j,k,l\in I}\widehat M_4(i,j,k,l)^2
$$
constitutes  a measure of deviation from Markovianness in terms of quadranucleotides analogously to $S_3$ for trinucleotides.

We repeated the experiments for trinucleotides shown in Tables~\ref{tab:3}
and~\ref{tab:4}, but using $S_4$ instead of $S_3$ as the measure of deviation
from Markovianness.  The results on the primary strand are displayed in
Table~\ref{tab:5} while those on the complementary strand appear in
Table~\ref{tab:6}.

\section{Further structure of the Markov chain}
\label{sec:further.struct}

\subsection{Markov chain with partitioned transition matrices}

Let $I$ be the set of START and STOP codon symbols and $Q=(Q_{ij}: i,j\in I)$ be
 the stochastic matrix of the Markov chain generating the sequence of annotated
START and STOP codons. Its stationary vector will be denoted by $\pi=(\pi_i:
i\in I)$. The entropy $h(Q,\pi)$ of this stationary Markov chain is
\begin{equation}
\label{entx1}
h(Q,\pi)=-\sum_{(i,j)\in I\times I} \pi_i q_{ij}\log q_{ij}.
\end{equation}

The set $I$ is partitioned into two disjoint sets: the set of START codons $I_0$ and the set of STOP codons $I_1$. 
Then, $Q$ has the form
\begin{equation}
\label{eq0}
Q=
\left(\begin{array}{cc}
0 & Q^0 \\
Q^1 & 0
\end{array}\right)
\end{equation}
That is $q_{ij}=0$ if $\{i,j\}\subseteq I_0$ or if $\{i,j\}\subseteq I_1$. the
matrix $Q^0$ is of dimension $|I_0|\times|I_1|$ while $Q^1$ is a
$|I_1|\times|I_0|$ matrix.  It is convenient to set
$$
\Delta=I_0\times I_1 \cup I_1\times I_0.
$$
The Markov measure $\PP(X_k=i_k, k=0,..,m)=
\pi_{i_k}\prod_{k=0}^{m-1} q_{i_k i_{k+1}}$
can give positive weight only to those trajectories with
$(i_k,i_{k+1})\in \Delta$ for all $k=0,..,m-1$.
In this case the entropy formula satisfies
$$
h(Q,\pi)=-\sum_{(i,j)\in \Delta}\pi_i q_{ij}\log q_{ij}.
$$
Since $Q$ is stochastic, the row sums of the matrices 
$Q^0$ and $Q^1$ are equal to $1$. Let ${\bf 1}_{I_l}$ be a  unitary
vector of dimension $|I_l|$ for $l=0,1$. Then, $Q^0{\bf 1}_{I_1}={\bf 1}_{I_0}$ and $Q^1{\bf 1}_{I_0}={\bf 1}_{I_1}$.
We can assume that the matrices $Q^1 Q^0$ and $Q^0 Q^1$ are strictly positive, 
which ensures that $Q$ is irreducible. Let 
$\pi$ be the unique stationary vector, we denote 
$\pi_{I_l}=(\pi_i: i\in I_l)$ for $l=0,1$. 
The stationary condition is $\pi^t Q=\pi^t$, 
which is equivalent to
$\pi_{I_0}^t=\pi_{I_1}^t Q^1$  and $\pi_{I_1}^t=\pi_{I_0}^t Q^0$.
Hence,
\begin{equation}
\label{eq2}
\pi_{I_0}^t=\pi_{I_0}^t Q^0 Q^1 \hbox{ and } \pi_{I_1}^t=\pi_{I_1}^t Q^1 Q^0. 
\end{equation}
The strictly positive matrices $Q^1 Q^0$ and $Q^0 Q^1$ are stochastic, so there exist positive solutions $\pi_{I_0}$ and $\pi_{I_1}$
to (\ref{eq2}) and we require two normalization conditions. 
The first one is $\pi_{I_0}^t {\bf 1}_{I_0}+\pi_{I_1}^t {\bf 1}_{I_1}=1$, that is 
$\sum_{i\in I_0}\pi_i+\sum_{i\in I_1}\pi_i=1$. The second condition is 
$$
\pi_{I_0}^t{\bf 1}_{I_0}=\pi_{I_1}^t Q^1{\bf 1}_{I_0}=\pi_{I_1}^t{\bf 1}_{I_1}.
$$ 
That is $\sum_{i\in I_0}\pi_i=\sum_{i\in I_1}\pi_i$, 
and so it is equal to $1/2$. 
Hence the probability vector $\pi$ is uniquely determined. 

\subsection{Conditional Independence}

It will be useful to set $I_l=I_0$ when $l$ is even and 
$I_l=I_1$ when $l$ is odd. In the sequel let $l$ be $0$ or $1$. 
The sequence $(X_n: n\ge 0)$ is conditionally independent given
$X_0\in I_l$ if and only if for all $m\ge 0$ and all 
$i_k\in I_{l+k}$, $k=0,..,m$, we have
\begin{equation}
\label{eq4}
\PP(X_k=i_k, k=0,..,m | X_0\in I_l)= \prod_{k=0}^m \PP(X_k=i_k | X_0\in I_l).
\end{equation}
This equality is easily seen to be equivalent to
$$
2\pi_{i_l}\prod_{k=0}^{m-1}q_{i_{k+l}i_{k+1+l}}=2^{m+1}\prod_{k=0}^m \pi_{i_k}.
$$
From this relation it can be concluded that 
a necessary and sufficient condition for conditional independence is 
$$
\forall\, i,j\in I:\quad q_{ij}=2\pi_j \ind_{(i,j)\in \Delta}.
$$ 

When there is conditional independence, and to avoid any confusion, 
the transition matrix will be denoted $Q^\dag=(q^\dag_{ij}:i,j\in I)$,
so $q^\dag_{ij}=2\pi_j \ind_{(i,j)\in \Delta}$.
In this case the invariant distribution is also $\pi$. 
We have
\begin{equation}
\label{entx2}
h(Q^\dag,\pi)=-\sum_{(i,j)\in \Delta} 2\pi_i \pi_j\log 2\pi_j=-\log 2+h_\pi,
\end{equation}
where we noted $h_\pi=-\sum_{i\in I} \pi_i \log \pi_i$
and we used $\sum_{i\in I_l} \pi_i=1/2$ and  $\sum_{(i,j)\in \Delta} 
\pi_i\pi_j=1/2$. 

\medskip

\noindent\textbf{Remark}.
In a similar way, we could also define conditional independence given the event $X_k\in I_{l+k}$, $k=0,..,s$, but the equality
$$
\PP(X_k=i_k, k=0,..,m | X_k\in I_{l+k}, k=0,..,s)
=\PP(X_k=i_k, k=0,..,m | X_0\in I_l)
$$
for $m\geq s$ means that this definition is equivalent to conditional independence given $X_0\in I_l$.

\subsection{Kullback-Leibler divergence and mutual information 
in the conditional case}

Let $\bP$ be the joint distribution of $(X_k,X_{k+1})$ on $I\times I$ 
for some $k\geq0$. By stationarity it does not depend on $k$. We write
$\bP(i,j)=\PP((X_k,X_{k+1}=(i,j))$ for $(i,j)\in I\times I$. 
Note that
$$
\bP(i,j)=\pi_i q_{ij} \ind_{(i,j)\in \Delta}. 
$$
So, $\bP$ is supported by $\Delta$.
The entropy of the measure $\bP$ on $I\times I$ is:
\begin{equation}
\label{entx3}
h(\bP)=-\sum_{(i,j)\in I\times I} \pi_i q_{ij} \log(\pi_i q_{ij})
=h(Q,\pi)+h_\pi.
\end{equation}
Let us consider $\bP^\dag$ as the bivariate distribution
under conditional independence. Then,
$$
\bP^\dag(i,j)=\pi_i q^\dag_{ij} \ind_{(i,j)\in \Delta}=
2\pi_i\pi_j\ind_{(i,j)\in \Delta}.
$$
The entropy of the joint distribution $\bP^\dag$ is
$$
h(\bP^\dag)= - \sum_{(i,j)\in \Delta} 2\pi_i\pi_j\log(2\pi_i\pi_j)
=-\log2 - 2\sum_{i\in I}\pi_i\log\pi_i =-\log2 +2h_\pi.
$$

Let us consider the Kullback-Leibler divergence of $\bP^\dag$ from~$\bP$. 
By definition, the divergence is
$$
\kld{\bP}{\bP^\dag}
= \sum_{i\in I,j\in I}\bP(i,j)\log(\bP(i,j)/\bP^\dag(i,j)).
$$
In our case an easy computation shows that 
$$
\kld{\bP}{\bP^\dag}=-\log2 +2h_\pi- h(\bP).
$$
These equalities, together with formulae (\ref{entx1}), 
(\ref{entx2}) and (\ref{entx3}) give
$$
\kld{\bP}{\bP^\dag}= h(Q^\dag,\pi) - h(Q,\pi).
$$
As $\bP$ and $\bP^\dag$ are proper probability distributions, 
Gibbs' inequality yields $\kld{\bP}{\bP^\dag}\geq 0$. Since we assume
$Q^0$ and $Q^1$ are strictly positive matrices, the 
distributions $\bP$ and 
$\bP^\dag$ have the same support $\Delta$. Then, 
we have that the inequality
$\kld{\bP}{\bP^\dag}\geq 0$ becomes a strict equality 
$\kld{\bP}{\bP^\dag}= 0$ if and only if $\bP=\bP^\dag$. Consequently, 
$\kld{\bP}{\bP^\dag}= h(Q^\dag,\pi) - h(Q,\pi)$ 
provides us a way to measure how closely $\PP$ complies with the notion 
of conditional independence described above.

\medskip

We can interpret the above result in terms of mutual information.
Let $\pi\otimes \pi$ be the product probability measure on $I\times I$.
The mutual information of the distribution ~$\bP$ of $(X_k,X_{k+1})$ on $I\times I$  
satisfies:
$$
I(\bP)=\sum_{(i,j)\in I\times I} \bP(i,j)
\log\frac{\bP(i,j)}{\pi\otimes \pi(i,j)}
=\sum_{(i,j)\in \Delta} \pi_i q_{ij}
\log\frac{\pi_i q_{ij}}{\pi_i\pi_j}=h_\pi - h(Q,\pi). 
$$
It follows that
$$
\kld{\bP}{\bP^\dag}= I(\bP)-\log2.
$$
Therefore, the mutual information is bounded below by 
$\log2$. Further, attainment 
of this lower bound by $I(\bP)$ is equivalent to 
$\kld{\bP}{\bP^\dag}=0$, that is, conditional independence of the
sequence $(X_n:n\in\NN)$ given $X_0\in I_l$ for some $l\in\{0,1\}$.  
Indeed, $\log2$ is the mutual information of conditionally 
independent random variables:
$$
I(\bP^\dag)
= \sum_{(i,j)\in\Delta} 2\pi_i\pi_j\log\frac{2\pi_i\pi_j}{\pi_i\pi_j}
=\sum_{(i,j)\in\Delta} 2\pi_i\pi_j\log2 = \log2.
$$
Thus, $\kld{\bP}{\bP^\dag} = I(\bP)-I(\bP^\dag)$.  The divergence
of~$\bP^\dag$ from~$\bP$ may thus be viewed as the difference in mutual
information of $\bP$ and $\bP^\dag$.

\medskip
\paragraph{\bf Remark.}

To consider the case in which there are more than two classes of codons,
let $I=I_0\cup I_1\cup\cdots\cup I_{d-1}$ where $d>1$.  Suppose that $Q_k$, $k=0,1,\ldots,d-1$ is a collection of stochastic matrices and that $Q$ has the form
$$
Q= \begin{pmatrix}%\left\{\begin{array}{ccccc}
0 & Q^0 & 0 & \cdots & 0 \\
0 & 0 & Q^1 & \cdots & 0 \\
\vdots & \vdots & \vdots & & \vdots \\
0 & 0 & 0 & \cdots & Q^{d-2} \\
Q^{d-1} & 0 & 0 & \cdots & 0
\end{pmatrix}.%\end{array}\right\}.
$$
then the above discussion remains valid for~$Q$ by replacing $\log2$ by $\log d$ and $\Delta$ by
$$
\Delta = \bigcup_{l=0}^{d-1} I_l\times I_{l+1} \times \cdots \times I_{l+d-1}
$$
where we set $I_{s+rd}=I_s$ for $s=0,\ldots,d-1$, $r\geq0$.
\QED

\bigskip

\begin{table}[ht]
\caption{Entropies, Kullback-Leibler divergences and relative difference in
entropies for the Markov chain producing the sequence of START and STOP codons
on the \textit{primary} strand of 13 bacterial chromosomes.}
\label{tab:7}
\bigskip
\centering
\begin{tabular}{p{8cm}rrrr}
  \hline
 Chromosome & Entropy & Entropy & K-L Div & Rel. Diff. \\
  & $h(Q,\pi)$ & $h(Q^\dag,\pi)$ & & (\%) \\
  \hline
Escherichia coli str. K-12 substr. MG1655 
& 0.5987 & 0.5991 & 0.0004 & 0.04 \\ 
  Helicobacter pylori 26695 chromosome 
& 0.7986 & 0.7994 & 0.0008 & 0.05 \\ 
  Staphylococcus aureus subsp. aureus MRSA252 chromosome
& 0.6738 & 0.6751 & 0.0013 & 0.10 \\ 
  Leptospira interrogans serovar Lai str. 56601 chromosome I 
& 0.8084 & 0.8104 & 0.0020 & 0.13 \\ 
  Leptospira interrogans serovar Lai str. 56601 chromosome II 
& 0.8107 & 0.8224 & 0.0116 & 0.71 \\ 
  Streptococcus pneumoniae ATCC 700669,  
& 0.6317 & 0.6323 & 0.0006 & 0.05 \\ 
  Bacillus subtilis subsp. spizizenii str. W23 chromosome
& 0.7957 & 0.7968 & 0.0011 & 0.07 \\ 
  Vibrio cholerae O1 str. 2010EL-1786 chromosome 1
& 0.7323 & 0.7350 & 0.0027 & 0.19 \\ 
  Vibrio cholerae O1 str. 2010EL-1786 chromosome 2
& 0.7473 & 0.7554 & 0.0081 & 0.54 \\ 
  Propionibacterium acnes TypeIA2 P.acn33 chromosome
& 0.6375 & 0.6424 & 0.0049 & 0.38 \\ 
  Salmonella enterica subsp. enterica serovar Typhi str. P-stx-12 
& 0.7226 & 0.7237 & 0.0012 & 0.08 \\ 
  Yersinia pestis D182038 chromosome 
& 0.7679 & 0.7695 & 0.0016 & 0.10 \\ 
  Mycobacterium tuberculosis 7199-99 
& 0.9113 & 0.9130 & 0.0017 & 0.09 \\ 
   \hline
\end{tabular}
\end{table}

\begin{table}[ht]
\caption{Entropies, Kullback-Leibler divergences and relative difference 
in entropies for the Markov chain producing the sequence of START and 
STOP codons on the \textit{complementary} strand of 13 bacterial chromosomes.}
\label{tab:8}
\bigskip
\centering
\begin{tabular}{p{8cm}rrrr}
  \hline
 Chromosome & Entropy & Entropy & K-L Div & Rel. Diff. \\
  & $h(Q,\pi)$ & $h(Q^\dag,\pi)$ & & (\%) \\
  \hline
Escherichia coli str. K-12 substr. MG1655 
& 0.5801 & 0.5816 & 0.0015 & 0.13 \\ 
  Helicobacter pylori 26695 chromosome. 
& 0.7734 & 0.7771 & 0.0037 & 0.24 \\ 
  Staphylococcus aureus subsp. aureus MRSA252 chromosome
& 0.6597 & 0.6611 & 0.0014 & 0.11 \\ 
  Leptospira interrogans serovar Lai str. 56601 chromosome I 
& 0.8187 & 0.8200 & 0.0013 & 0.08 \\ 
  Leptospira interrogans serovar Lai str. 56601 chromosome II 
& 0.8078 & 0.8350 & 0.0272 & 1.66 \\ 
  Streptococcus pneumoniae ATCC 700669 
& 0.6462 & 0.6491 & 0.0029 & 0.22 \\ 
  Bacillus subtilis subsp. spizizenii str. W23 chromosome
& 0.7878 & 0.7885 & 0.0007 & 0.04 \\ 
  Vibrio cholerae O1 str. 2010EL-1786 chromosome 1
& 0.7320 & 0.7343 & 0.0023 & 0.16 \\ 
  Vibrio cholerae O1 str. 2010EL-1786 chromosome 2
& 0.7491 & 0.7634 & 0.0143 & 0.95 \\ 
  Propionibacterium acnes TypeIA2 P.acn33 chromosome
& 0.6474 & 0.6493 & 0.0019 & 0.15 \\ 
  Salmonella enterica subsp. enterica serovar Typhi str. P-stx-12 
& 0.7272 & 0.7276 & 0.0004 & 0.03 \\ 
  Yersinia pestis D182038 
\newline  chromosome
& 0.7607 & 0.7641 & 0.0034 & 0.22 \\ 
  Mycobacterium tuberculosis 7199-99 
& 0.9052 & 0.9071 & 0.0019 & 0.10 \\ 
\hline
\end{tabular}
\end{table}

For both strands, we calculated the entropies $h(Q,\pi)$ and $h(Q^\dag,\pi)$, as
 l as the Kullback-Leibler divergence and the relative difference between the
entropies expressed as a percentage.  All of these values a summarized in
Tables~\ref{tab:7} and~\ref{tab:8}.  The Kullback-Lebler divergences and the
relative differences between the entropies are all very close to zero, which is
precisely what one expects to see if $(X_n:n\in\NN)$ is conditionally independent
given $X_0\in I_l$ for some $l\in\{0,1\}$.

We need to mention that the transition matrices for escherichia coli K-12
violate the exact form of~$Q$ (see the comments in the first section of the
appendix).  They possess some non-zero elements in the top-left quadrant of the
matrix.  In order to calculate the quantities shown in Tables~\ref{tab:7}
and~\ref{tab:8}, it was necessary to set the offending elements to zero and
rescale the affected rows to sum to unity.

\begin{table}[tbp]
\caption{Maximum absolute difference between START/STOP frequencies on the two
strands in the DNA duplexes of 13 bacterial chromosomes.  the last column displays the number of annotated START/STOP codons present in the duplex.}
\label{tab:9}
\bigskip
\centering
\begin{tabular}{p{8cm}rrr}
  \hline
 Chromosome & Maximum Absolute & Corr. & Number of \\ 
  & Difference in Frequency & Coef. & Codons \\
  \hline
Escherichia coli str. K-12 substr. MG1655 & 0.0095 & 0.9995 & 8342 \\ 
  Helicobacter pylori 26695 chromosome & 0.0104 & 0.9994 & 3134 \\ 
  Staphylococcus aureus subsp. aureus MRSA252 chromosome & 0.0123 & 0.9988 & 5290 \\ 
  Leptospira interrogans serovar Lai str. 56601 chromosome I & 0.0093 & 0.9993 & 6818 \\ 
  Leptospira interrogans serovar Lai str. 56601 chromosome II & 0.0240 & 0.9932 & 586 \\ 
  Streptococcus pneumoniae ATCC 700669 & 0.0189 & 0.9983 & 3980 \\ 
  Bacillus subtilis subsp. spizizenii str. W23 chromosome & 0.0083 & 0.9991 & 8120 \\ 
  Vibrio cholerae O1 str. 2010EL-1786 chromosome 1 & 0.0122 & 0.9985 & 5610 \\ 
  Vibrio cholerae O1 str. 2010EL-1786 chromosome 2 & 0.0420 & 0.9825 & 2080 \\ 
  Propionibacterium acnes TypeIA2 P.acn33 chromosome & 0.0046 & 0.9999 & 4466 \\ 
  Salmonella enterica subsp. enterica serovar Typhi str. P-stx-12 & 0.0219 & 0.9961 & 9380 \\ 
  Yersinia pestis D182038 chromosome & 0.0044 & 0.9998 & 7240 \\ 
  Mycobacterium tuberculosis 7199-99 & 0.0069 & 0.9993 & 7968 \\ 
\hline
\end{tabular}
\end{table}

\subsection{Chargaff's second parity rule}

Finally, we have observed that the annotated START and STOP codons taken together from
both the primary and complementary strands essentially comply with Chargaff's
second parity rule.  Chargaff's first parity rule~\cite{chargaff} says that,
within a DNA duplex, the numbers of~$A$ and~$T$ mononucleotides are the same
while the numbers of~$C$ and~$G$ nucleotides also agree. Chargaff's second
parity rule not only says this continues to hold within a DNA simplex, but
rather that short oligonucleotides and their reverse complements appear with the
 same frequency within a simplex~\cite{chargaff2}.  Firstly, note that within a
DNA duplex, the START and STOP codons on one strand correspond to their reverse
complements on the other strand.  Consequently, we
can perform basic checks for compliance with Chargaff's second parity rule by
considering both the difference and correlation between the frequencies of every START and STOP codon
 in each strand.  If we let $\pi^{(1)}=(\pi_i^{(1)}: i\in I)$ and
$\pi^{(2)}=(\pi_i^{(2)}: i\in I)$ be the frequencies of the symbols in~$I$ on
the primary and complementary strands respectively.  Of course, $\pi^{(1)}$ and
$\pi^{(2)}$ constitute the stationary distributions of the chains of START and
STOP codons on their corresponding strands.  The $\ell_\infty$ distance between
these two probability vectors, given by
$$
\norm{\pi^{(1)}-\pi^{(2)}}_\infty
= \max_{i\in I}\abs{\pi_i^{(1)}-\pi_i^{(2)}},
$$
together with their sample correlation coefficient $\corr\left(\pi^{(1)},
\pi^{(2)}\right)$, will indicate how closely the trinucleotide frequencies
conform to Chargaff's second parity rule.

Table~\ref{tab:9} shows $\norm{\pi^{(1)}- \pi^{(2)}}_\infty$ and
$\corr\left(\pi^{(1)}, \pi^{(2)}\right)$ for each of the 13 DNA sequences we
have examined.  the values shown in the table indicate a high degree of
compliance by the START/STOP sequences with Chargaff's second parity rule.  The
last column of the table shows the number of codons in the DNA duplex of each
chromosome. The annotated START and STOP codons in the bacterial duplexes
examined constitute between 1578 and 28140 nucleotides with a mean average of
16848. Generally speaking, this is equivalent to discovering Chargaff's second
parity rule in short sequences, but the level of compliance based on the
correlation (0.9825--0.9999)we have observed for START/STOP codon sequences
appears high for the quantity of nucleotides.  It is instructive to compare this
 to that reported in Figure~4a of~\cite{PNAS_Albrecht} for nucleotide sequence
segments of comparable size taken from human chromosome 1, but with two caveats.
 Firstly, note  that the correlations reported in Table~\ref{tab:9} are based on
 the vectors $\pi^{(1)}$ and $\pi^{(2)}$, which are of length~6 or~7 for the
bacteria studied here, whereas Albrecht-Buehler's correlations are based on
vectors containing the counts for 64 trinucleotides.  This may partly account
for the high levels of compliance and small variance seen here, even for very
short codon sequences. Secondly, we are comparing intrastrand codon correlations
 or prokaryotes against those for a eukaryote chromosome, which strictly
speaking should not be comparable since they may respond in different ways to
varying quantities of nucleotides.

\section*{Acknowledgements}

This work was supported by the Center for Mathematical Modeling  (CMM) Basal
CONICYT Program PFB 03 and INRIA-CIRIC-CHILE Program Natural Resources.
The authors also extend their thanks to the Laboratory of Bioinformatics and
Mathematics of the Genome (LBMG) for invaluable assistance.

%\bibliographystyle{plain}
%\bibliography{../bib/DNAProb,../bib/test,../bib/dna,../bib/stats}

\begin{thebibliography}{10}

\bibitem{PNAS_Albrecht}
G.~Albrecht-Buehler.
\newblock Asymptotically increasing compliance of genomes with {Chargaff's}
  second parity rules through inversions and inverted transpositions.
\newblock {\em PNAS}, 103(47):17828--17833, 2006.

\bibitem{bouaynaya&schonfeld2008}
N.~Bouaynaya and D.~Schonfeld.
\newblock Non-stationary analysis of coding and non-coding regions in
  nucleotide sequences.
\newblock {\em IEEE J. of Selected Topics in Signal Processing}, 2(3):357--364,
  2008.

\bibitem{chargaff}
E.~Chargaff.
\newblock Chemical specificity of nucleic acids and mechanism of their
  enzymatic degradation.
\newblock {\em Experientia}, 6(6):201--9, 1950.

\bibitem{degroot1991}
M.H. DeGroot.
\newblock {\em Probability and Statistics}.
\newblock Addison-Wesley, Reading, MA, 3rd edition, 1991.

\bibitem{Fedorov&etal2002}
A.~Fedorov, S.~Saxonov, and W.~Gilbert.
\newblock Regularities of context-dependent codon bias in eukaryotic genes.
\newblock {\em Nucleic Acids Res.}, 30(5):1192--1197, 2002.

\bibitem{hart&martinez2011}
A.G. Hart and S.~Mart{\'\i}nez.
\newblock Statistical testing of {Chargaff's} second parity rule in bacterial
  genome sequences.
\newblock {\em Stoch. Models}, 27(2):1--46, 2011.

\bibitem{li&kaneko1992}
W.~Li and Kaneko. K.
\newblock Long-range correlation and partial 1/f spectrum in a noncoding dna
  sequence.
\newblock {\em Europhysics Letters}, 17:655, 1992.

\bibitem{ljung&box1976}
G.M. Ljung and G.E.P. Box.
\newblock On a measure of lack of fit in time series models.
\newblock {\em Biometrika}, 65(2):297--303, 1976.

\bibitem{peng&etal1992}
C.~K. Peng, S.~V. Buldyrev, A.~L. Goldberger, S.~Havlin, F.~Sciortino, H.~M.
  Simons, and E.~Stanley.
\newblock Long-range correlations in nucleotide sequences.
\newblock {\em Nature}, 356(6365):168--170, 1992.

\bibitem{chargaff2}
R.~Rudner, J.D. Karkas, and E.~Chargaff.
\newblock Separation of {B}. subtilis {DNA} into complementary strands. {III}.
  direct analysis.
\newblock {\em Proc Natl Acad Sci USA}, 60:921--922, 1968.

\end{thebibliography}

\newpage
\appendix
\section*{Appendix:  Estimated transition matrices for 13 bacterial genomes}

%matrices.tex

\begin{enumerate}
\item Escherichia coli str. K-12 substr. MG1655, complete genome.

In the estimated START/STOP transition matrices for the primary 
and complementary strands of Escherichia Coli K-12, there appear 
two anomalous entries in the top-left corner of each matrix.  
Inspection of the annotation (NC\_000913.2) available from GenBank 
reveals that the 603rd gene on the primary strand spans loci 
1204594--1205365 relative to the $5'$ end
it starts with GTG and finishes with an ATG codon.

Similarly, the two non-zero elements in the top-left corner of the 
transition matrix estimated for the complementary strand are 
explained by the 473rd gene on the complementary strand.  
This gene spans loci 1077648--1077866 relative to the $5'$ 
end of the complementary strand.  It starts with an
ATG codon and finishes with a GTG codon.

\begin{description}
\item[Primary Strand]

\begin{center}
\begin{tabular}{lrrrrrr}
 & ATG & GTG & TTG & TAA & TAG & TGA \\
ATG & 0.0005 & 0.0000 & 0.0000 & 0.6481 & 0.0784 & 0.2729 \\
  GTG & 0.0063 & 0.0000 & 0.0000 & 0.6266 & 0.0823 & 0.2848 \\
  TTG & 0.0000 & 0.0000 & 0.0000 & 0.6389 & 0.0833 & 0.2778 \\
  TAA & 0.8994 & 0.0831 & 0.0175 & 0.0000 & 0.0000 & 0.0000 \\
  TAG & 0.8875 & 0.0938 & 0.0187 & 0.0000 & 0.0000 & 0.0000 \\
  TGA & 0.9209 & 0.0612 & 0.0180 & 0.0000 & 0.0000 & 0.0000 \\
\end{tabular}
\end{center}

\item[Complementary Strand]

\begin{center}
\begin{tabular}{lrrrrrr}
 & ATG & GTG & TTG & TAA & TAG & TGA \\
ATG & 0.0000 & 0.0005 & 0.0000 & 0.6572 & 0.0556 & 0.2867 \\
  GTG & 0.0061 & 0.0000 & 0.0000 & 0.5521 & 0.0982 & 0.3436 \\
  TTG & 0.0000 & 0.0000 & 0.0000 & 0.5405 & 0.1081 & 0.3514 \\
  TAA & 0.9135 & 0.0707 & 0.0159 & 0.0000 & 0.0000 & 0.0000 \\
  TAG & 0.8984 & 0.0781 & 0.0234 & 0.0000 & 0.0000 & 0.0000 \\
  TGA & 0.8946 & 0.0863 & 0.0192 & 0.0000 & 0.0000 & 0.0000 \\
\end{tabular}
\end{center}
\end{description}

\item Helicobacter pylori 26695 chromosome, complete genome.

\begin{description}
\item[Primary Strand]

\begin{center}
\begin{tabular}{lrrrrrr}
 & ATG & GTG & TTG & TAA & TAG & TGA \\
ATG & 0.0000 & 0.0000 & 0.0000 & 0.5563 & 0.1688 & 0.2749 \\
  GTG & 0.0000 & 0.0000 & 0.0000 & 0.4750 & 0.2000 & 0.3250 \\
  TTG & 0.0000 & 0.0000 & 0.0000 & 0.5323 & 0.1935 & 0.2742 \\
  TAA & 0.8106 & 0.1103 & 0.0791 & 0.0000 & 0.0000 & 0.0000 \\
  TAG & 0.8120 & 0.0977 & 0.0902 & 0.0000 & 0.0000 & 0.0000 \\
  TGA & 0.8224 & 0.0981 & 0.0794 & 0.0000 & 0.0000 & 0.0000 \\
\end{tabular}
\end{center}

\item[Complementary Strand]

\begin{center}
\begin{tabular}{lrrrrrr}
 & ATG & GTG & TTG & TAA & TAG & TGA \\
ATG & 0.0000 & 0.0000 & 0.0000 & 0.5852 & 0.1493 & 0.2655 \\
  GTG & 0.0000 & 0.0000 & 0.0000 & 0.4583 & 0.1944 & 0.3472 \\
  TTG & 0.0000 & 0.0000 & 0.0000 & 0.5000 & 0.2500 & 0.2500 \\
  TAA & 0.8374 & 0.0879 & 0.0747 & 0.0000 & 0.0000 & 0.0000 \\
  TAG & 0.7692 & 0.1077 & 0.1231 & 0.0000 & 0.0000 & 0.0000 \\
  TGA & 0.8349 & 0.0826 & 0.0826 & 0.0000 & 0.0000 & 0.0000 \\
\end{tabular}
\end{center}
\end{description}

\item Staphylococcus aureus subsp. aureus MRSA252 chromosome, complete

\begin{description}
\item[Primary Strand]

\begin{center}
\begin{tabular}{lrrrrrr}
 & ATG & GTG & TTG & TAA & TAG & TGA \\
ATG & 0.0000 & 0.0000 & 0.0000 & 0.7486 & 0.1434 & 0.1080 \\
  GTG & 0.0000 & 0.0000 & 0.0000 & 0.7184 & 0.1748 & 0.1068 \\
  TTG & 0.0000 & 0.0000 & 0.0000 & 0.7023 & 0.1221 & 0.1756 \\
  TAA & 0.8219 & 0.0780 & 0.1001 & 0.0000 & 0.0000 & 0.0000 \\
  TAG & 0.7880 & 0.0924 & 0.1196 & 0.0000 & 0.0000 & 0.0000 \\
  TGA & 0.8231 & 0.0816 & 0.0952 & 0.0000 & 0.0000 & 0.0000 \\
\end{tabular}
\end{center}

\item[Complementary Strand]

\begin{center}
\begin{tabular}{lrrrrrr}
 & ATG & GTG & TTG & TAA & TAG & TGA \\
ATG & 0.0000 & 0.0000 & 0.0000 & 0.7276 & 0.1601 & 0.1123 \\
  GTG & 0.0000 & 0.0000 & 0.0000 & 0.7767 & 0.1262 & 0.0971 \\
  TTG & 0.0000 & 0.0000 & 0.0000 & 0.6460 & 0.1858 & 0.1681 \\
  TAA & 0.8483 & 0.0748 & 0.0768 & 0.0000 & 0.0000 & 0.0000 \\
  TAG & 0.8165 & 0.0872 & 0.0963 & 0.0000 & 0.0000 & 0.0000 \\
  TGA & 0.8354 & 0.0633 & 0.1013 & 0.0000 & 0.0000 & 0.0000 \\
\end{tabular}
\end{center}
\end{description}

\item Leptospira interrogans serovar Lai str. 56601 chromosome I,

\begin{description}
\item[Primary Strand]

\begin{center}
\begin{tabular}{lrrrrrr}
 & ATG & GTG & TTG & TAA & TAG & TGA \\
ATG & 0.0000 & 0.0000 & 0.0000 & 0.5717 & 0.1278 & 0.3004 \\
  GTG & 0.0000 & 0.0000 & 0.0000 & 0.6371 & 0.1694 & 0.1935 \\
  TTG & 0.0000 & 0.0000 & 0.0000 & 0.5338 & 0.1673 & 0.2989 \\
  TAA & 0.7892 & 0.0609 & 0.1499 & 0.0000 & 0.0000 & 0.0000 \\
  TAG & 0.7500 & 0.0968 & 0.1532 & 0.0000 & 0.0000 & 0.0000 \\
  TGA & 0.7646 & 0.0697 & 0.1657 & 0.0000 & 0.0000 & 0.0000 \\
\end{tabular}
\end{center}

\item[Complementary Strand]

\begin{center}
\begin{tabular}{lrrrrrr}
 & ATG & GTG & TTG & TAA & TAG & TGA \\
ATG & 0.0000 & 0.0000 & 0.0000 & 0.5846 & 0.1230 & 0.2923 \\
  GTG & 0.0000 & 0.0000 & 0.0000 & 0.5000 & 0.1944 & 0.3056 \\
  TTG & 0.0000 & 0.0000 & 0.0000 & 0.5704 & 0.1480 & 0.2816 \\
  TAA & 0.7663 & 0.0598 & 0.1739 & 0.0000 & 0.0000 & 0.0000 \\
  TAG & 0.7299 & 0.0806 & 0.1896 & 0.0000 & 0.0000 & 0.0000 \\
  TGA & 0.7570 & 0.0774 & 0.1656 & 0.0000 & 0.0000 & 0.0000 \\
\end{tabular}
\end{center}
\end{description}

\item Leptospira interrogans serovar Lai str. 56601 chromosome II,

\begin{description}
\item[Primary Strand]

\begin{center}
\begin{tabular}{lrrrrrr}
 & ATG & GTG & TTG & TAA & TAG & TGA \\
ATG & 0.0000 & 0.0000 & 0.0000 & 0.5372 & 0.1405 & 0.3223 \\
  GTG & 0.0000 & 0.0000 & 0.0000 & 0.6000 & 0.0000 & 0.4000 \\
  TTG & 0.0000 & 0.0000 & 0.0000 & 0.6207 & 0.1034 & 0.2759 \\
  TAA & 0.7303 & 0.0562 & 0.2135 & 0.0000 & 0.0000 & 0.0000 \\
  TAG & 0.8500 & 0.1000 & 0.0500 & 0.0000 & 0.0000 & 0.0000 \\
  TGA & 0.7647 & 0.0588 & 0.1765 & 0.0000 & 0.0000 & 0.0000 \\
\end{tabular}
\end{center}

\item[Complementary Strand]

\begin{center}
\begin{tabular}{lrrrrrr}
 & ATG & GTG & TTG & TAA & TAG & TGA \\
ATG & 0.0000 & 0.0000 & 0.0000 & 0.6186 & 0.1237 & 0.2577 \\
  GTG & 0.0000 & 0.0000 & 0.0000 & 0.1111 & 0.2222 & 0.6667 \\
  TTG & 0.0000 & 0.0000 & 0.0000 & 0.6667 & 0.1481 & 0.1852 \\
  TAA & 0.7089 & 0.0506 & 0.2405 & 0.0000 & 0.0000 & 0.0000 \\
  TAG & 0.6667 & 0.1111 & 0.2222 & 0.0000 & 0.0000 & 0.0000 \\
  TGA & 0.8056 & 0.0833 & 0.1111 & 0.0000 & 0.0000 & 0.0000 \\
\end{tabular}
\end{center}
\end{description}

\item Streptococcus pneumoniae ATCC 700669, complete genome.

\begin{description}
\item[Primary Strand]

\begin{center}
\begin{tabular}{lrrrrrr}
 & ATG & GTG & TTG & TAA & TAG & TGA \\
ATG & 0.0000 & 0.0000 & 0.0000 & 0.6375 & 0.2071 & 0.1554 \\
  GTG & 0.0000 & 0.0000 & 0.0000 & 0.6667 & 0.2308 & 0.1026 \\
  TTG & 0.0000 & 0.0000 & 0.0000 & 0.6383 & 0.2340 & 0.1277 \\
  TAA & 0.9049 & 0.0443 & 0.0508 & 0.0000 & 0.0000 & 0.0000 \\
  TAG & 0.9200 & 0.0300 & 0.0500 & 0.0000 & 0.0000 & 0.0000 \\
  TGA & 0.9172 & 0.0414 & 0.0414 & 0.0000 & 0.0000 & 0.0000 \\
\end{tabular}
\end{center}

\item[Complementary Strand]

\begin{center}
\begin{tabular}{lrrrrrr}
 & ATG & GTG & TTG & TAA & TAG & TGA \\
ATG & 0.0000 & 0.0000 & 0.0000 & 0.5979 & 0.2349 & 0.1672 \\
  GTG & 0.0000 & 0.0000 & 0.0000 & 0.6000 & 0.2400 & 0.1600 \\
  TTG & 0.0000 & 0.0000 & 0.0000 & 0.6750 & 0.2250 & 0.1000 \\
  TAA & 0.9100 & 0.0418 & 0.0482 & 0.0000 & 0.0000 & 0.0000 \\
  TAG & 0.9012 & 0.0782 & 0.0206 & 0.0000 & 0.0000 & 0.0000 \\
  TGA & 0.9412 & 0.0294 & 0.0294 & 0.0000 & 0.0000 & 0.0000 \\
\end{tabular}
\end{center}
\end{description}

\item Bacillus subtilis subsp. spizizenii str. W23 chromosome, complete

\begin{description}
\item[Primary Strand]

\begin{center}
\begin{tabular}{lrrrrrr}
 & ATG & GTG & TTG & TAA & TAG & TGA \\
ATG & 0.0000 & 0.0000 & 0.0000 & 0.6508 & 0.1247 & 0.2244 \\
  GTG & 0.0000 & 0.0000 & 0.0000 & 0.6051 & 0.1385 & 0.2564 \\
  TTG & 0.0000 & 0.0000 & 0.0000 & 0.5714 & 0.1389 & 0.2897 \\
  TAA & 0.7602 & 0.1047 & 0.1350 & 0.0000 & 0.0000 & 0.0000 \\
  TAG & 0.7683 & 0.1057 & 0.1260 & 0.0000 & 0.0000 & 0.0000 \\
  TGA & 0.7863 & 0.0903 & 0.1233 & 0.0000 & 0.0000 & 0.0000 \\
\end{tabular}
\end{center}

\item[Complementary Strand]

\begin{center}
\begin{tabular}{lrrrrrr}
 & ATG & GTG & TTG & TAA & TAG & TGA \\
ATG & 0.0000 & 0.0000 & 0.0000 & 0.6388 & 0.1370 & 0.2243 \\
  GTG & 0.0000 & 0.0000 & 0.0000 & 0.5870 & 0.1902 & 0.2228 \\
  TTG & 0.0000 & 0.0000 & 0.0000 & 0.6064 & 0.1596 & 0.2340 \\
  TAA & 0.7780 & 0.0913 & 0.1307 & 0.0000 & 0.0000 & 0.0000 \\
  TAG & 0.7864 & 0.0809 & 0.1327 & 0.0000 & 0.0000 & 0.0000 \\
  TGA & 0.7905 & 0.0747 & 0.1349 & 0.0000 & 0.0000 & 0.0000 \\
\end{tabular}
\end{center}
\end{description}

\item Vibrio cholerae O1 str. 2010EL-1786 chromosome 1, complete

\begin{description}
\item[Primary Strand]

\begin{center}
\begin{tabular}{lrrrrrr}
 & ATG & GTG & TTG & TAA & TAG & TGA \\
ATG & 0.0000 & 0.0000 & 0.0000 & 0.6291 & 0.1627 & 0.2083 \\
  GTG & 0.0000 & 0.0000 & 0.0000 & 0.4963 & 0.2296 & 0.2741 \\
  TTG & 0.0000 & 0.0000 & 0.0000 & 0.5128 & 0.2308 & 0.2564 \\
  TAA & 0.8508 & 0.0955 & 0.0537 & 0.0000 & 0.0000 & 0.0000 \\
  TAG & 0.8109 & 0.1134 & 0.0756 & 0.0000 & 0.0000 & 0.0000 \\
  TGA & 0.8562 & 0.0936 & 0.0502 & 0.0000 & 0.0000 & 0.0000 \\
\end{tabular}
\end{center}

\item[Complementary Strand]

\begin{center}
\begin{tabular}{lrrrrrr}
 & ATG & GTG & TTG & TAA & TAG & TGA \\
ATG & 0.0000 & 0.0000 & 0.0000 & 0.6472 & 0.1689 & 0.1839 \\
  GTG & 0.0000 & 0.0000 & 0.0000 & 0.6000 & 0.2154 & 0.1846 \\
  TTG & 0.0000 & 0.0000 & 0.0000 & 0.5192 & 0.1731 & 0.3077 \\
  TAA & 0.8377 & 0.0916 & 0.0706 & 0.0000 & 0.0000 & 0.0000 \\
  TAG & 0.8387 & 0.1008 & 0.0605 & 0.0000 & 0.0000 & 0.0000 \\
  TGA & 0.8297 & 0.0797 & 0.0906 & 0.0000 & 0.0000 & 0.0000 \\
\end{tabular}
\end{center}
\end{description}

\item Vibrio cholerae O1 str. 2010EL-1786 chromosome 2, complete

\begin{description}
\item[Primary Strand]

\begin{center}
\begin{tabular}{lrrrrrr}
 & ATG & GTG & TTG & TAA & TAG & TGA \\
ATG & 0.0000 & 0.0000 & 0.0000 & 0.6527 & 0.1715 & 0.1757 \\
  GTG & 0.0000 & 0.0000 & 0.0000 & 0.5789 & 0.2105 & 0.2105 \\
  TTG & 0.0000 & 0.0000 & 0.0000 & 0.3913 & 0.2609 & 0.3478 \\
  TAA & 0.8072 & 0.1019 & 0.0909 & 0.0000 & 0.0000 & 0.0000 \\
  TAG & 0.8491 & 0.0660 & 0.0849 & 0.0000 & 0.0000 & 0.0000 \\
  TGA & 0.8482 & 0.1161 & 0.0357 & 0.0000 & 0.0000 & 0.0000 \\
\end{tabular}
\end{center}

\item[Complementary Strand]

\begin{center}
\begin{tabular}{lrrrrrr}
 & ATG & GTG & TTG & TAA & TAG & TGA \\
ATG & 0.0000 & 0.0000 & 0.0000 & 0.5891 & 0.1680 & 0.2429 \\
  GTG & 0.0000 & 0.0000 & 0.0000 & 0.3913 & 0.2174 & 0.3913 \\
  TTG & 0.0000 & 0.0000 & 0.0000 & 0.2653 & 0.2449 & 0.4898 \\
  TAA & 0.8480 & 0.0640 & 0.0880 & 0.0000 & 0.0000 & 0.0000 \\
  TAG & 0.8537 & 0.0366 & 0.1098 & 0.0000 & 0.0000 & 0.0000 \\
  TGA & 0.8268 & 0.0315 & 0.1417 & 0.0000 & 0.0000 & 0.0000 \\
\end{tabular}
\end{center}
\end{description}

\item Propionibacterium acnes TypeIA2 P.acn33 chromosome, complete

\begin{description}
\item[Primary Strand]

\begin{center}
\begin{tabular}{lrrrrrr}
 & ATG & GTG & TTG & TAA & TAG & TGA \\
ATG & 0.0000 & 0.0000 & 0.0000 & 0.1058 & 0.0579 & 0.8363 \\
  GTG & 0.0000 & 0.0000 & 0.0000 & 0.1199 & 0.0959 & 0.7842 \\
  TTG & 0.0000 & 0.0000 & 0.0000 & 0.2581 & 0.0323 & 0.7097 \\
  TAA & 0.6850 & 0.2913 & 0.0236 & 0.0000 & 0.0000 & 0.0000 \\
  TAG & 0.5467 & 0.4000 & 0.0533 & 0.0000 & 0.0000 & 0.0000 \\
  TGA & 0.7279 & 0.2459 & 0.0262 & 0.0000 & 0.0000 & 0.0000 \\
\end{tabular}
\end{center}

\item[Complementary Strand]

\begin{center}
\begin{tabular}{lrrrrrr}
 & ATG & GTG & TTG & TAA & TAG & TGA \\
ATG & 0.0000 & 0.0000 & 0.0000 & 0.1098 & 0.0600 & 0.8301 \\
  GTG & 0.0000 & 0.0000 & 0.0000 & 0.1174 & 0.0772 & 0.8054 \\
  TTG & 0.0000 & 0.0000 & 0.0000 & 0.1143 & 0.1429 & 0.7429 \\
  TAA & 0.6880 & 0.2640 & 0.0480 & 0.0000 & 0.0000 & 0.0000 \\
  TAG & 0.6133 & 0.3333 & 0.0533 & 0.0000 & 0.0000 & 0.0000 \\
  TGA & 0.7107 & 0.2620 & 0.0273 & 0.0000 & 0.0000 & 0.0000 \\
\end{tabular}
\end{center}
\end{description}

\item Salmonella enterica subsp. enterica serovar Typhi str. P-stx-12

\begin{description}
\item[Primary Strand]

\begin{center}
\begin{tabular}{lrrrrrr}
 & ATG & GTG & TTG & TAA & TAG & TGA \\
ATG & 0.0000 & 0.0000 & 0.0000 & 0.5693 & 0.1013 & 0.3294 \\
  GTG & 0.0000 & 0.0000 & 0.0000 & 0.4891 & 0.1397 & 0.3712 \\
  TTG & 0.0000 & 0.0000 & 0.0000 & 0.5649 & 0.0992 & 0.3359 \\
  TAA & 0.8540 & 0.0860 & 0.0600 & 0.0000 & 0.0000 & 0.0000 \\
  TAG & 0.8413 & 0.1151 & 0.0437 & 0.0000 & 0.0000 & 0.0000 \\
  TGA & 0.8466 & 0.1047 & 0.0486 & 0.0000 & 0.0000 & 0.0000 \\
\end{tabular}
\end{center}

\item[Complementary Strand]

\begin{center}
\begin{tabular}{lrrrrrr}
 & ATG & GTG & TTG & TAA & TAG & TGA \\
ATG & 0.0000 & 0.0000 & 0.0000 & 0.6074 & 0.1017 & 0.2909 \\
  GTG & 0.0000 & 0.0000 & 0.0000 & 0.6144 & 0.0932 & 0.2924 \\
  TTG & 0.0000 & 0.0000 & 0.0000 & 0.5594 & 0.1189 & 0.3217 \\
  TAA & 0.8374 & 0.1055 & 0.0571 & 0.0000 & 0.0000 & 0.0000 \\
  TAG & 0.8283 & 0.0944 & 0.0773 & 0.0000 & 0.0000 & 0.0000 \\
  TGA & 0.8299 & 0.1015 & 0.0687 & 0.0000 & 0.0000 & 0.0000 \\
\end{tabular}
\end{center}
\end{description}

\item Yersinia pestis D182038 chromosome, complete genome.

\begin{description}
\item[Primary Strand]

\begin{center}
\begin{tabular}{lrrrrrr}
 & ATG & GTG & TTG & TAA & TAG & TGA \\
ATG & 0.0000 & 0.0000 & 0.0000 & 0.5631 & 0.1409 & 0.2960 \\
  GTG & 0.0000 & 0.0000 & 0.0000 & 0.5172 & 0.1724 & 0.3103 \\
  TTG & 0.0000 & 0.0000 & 0.0000 & 0.6230 & 0.1311 & 0.2459 \\
  TAA & 0.8383 & 0.0881 & 0.0736 & 0.0000 & 0.0000 & 0.0000 \\
  TAG & 0.7886 & 0.1220 & 0.0894 & 0.0000 & 0.0000 & 0.0000 \\
  TGA & 0.8254 & 0.1171 & 0.0575 & 0.0000 & 0.0000 & 0.0000 \\
\end{tabular}
\end{center}

\item[Complementary Strand]

\begin{center}
\begin{tabular}{lrrrrrr}
 & ATG & GTG & TTG & TAA & TAG & TGA \\
ATG & 0.0000 & 0.0000 & 0.0000 & 0.5433 & 0.1508 & 0.3059 \\
  GTG & 0.0000 & 0.0000 & 0.0000 & 0.5222 & 0.1556 & 0.3222 \\
  TTG & 0.0000 & 0.0000 & 0.0000 & 0.7218 & 0.0752 & 0.2030 \\
  TAA & 0.8379 & 0.0825 & 0.0796 & 0.0000 & 0.0000 & 0.0000 \\
  TAG & 0.8489 & 0.1043 & 0.0468 & 0.0000 & 0.0000 & 0.0000 \\
  TGA & 0.8252 & 0.1119 & 0.0629 & 0.0000 & 0.0000 & 0.0000 \\
\end{tabular}
\end{center}
\end{description}

\item Mycobacterium tuberculosis 7199-99 complete genome.

\begin{description}
\item[Primary Strand]

\begin{center}
\begin{tabular}{lrrrrrrr}
 & ATG & CTG & GTG & TTG & TAA & TAG & TGA \\
ATG & 0.0000 & 0.0000 & 0.0000 & 0.0000 & 0.1513 & 0.3059 & 0.5428 \\
  CTG & 0.0000 & 0.0000 & 0.0000 & 0.0000 & 0.0000 & 0.1429 & 0.8571 \\
  GTG & 0.0000 & 0.0000 & 0.0000 & 0.0000 & 0.1612 & 0.2701 & 0.5687 \\
  TTG & 0.0000 & 0.0000 & 0.0000 & 0.0000 & 0.2091 & 0.2455 & 0.5455 \\
  TAA & 0.6254 & 0.0032 & 0.3206 & 0.0508 & 0.0000 & 0.0000 & 0.0000 \\
  TAG & 0.5852 & 0.0069 & 0.3546 & 0.0534 & 0.0000 & 0.0000 & 0.0000 \\
  TGA & 0.6134 & 0.0018 & 0.3279 & 0.0569 & 0.0000 & 0.0000 & 0.0000 \\
\end{tabular}
\end{center}

\item[Complementary Strand]

\begin{center}
\begin{tabular}{lrrrrrrr}
 & ATG & CTG & GTG & TTG & TAA & TAG & TGA \\
ATG & 0.0000 & 0.0000 & 0.0000 & 0.0000 & 0.1672 & 0.3045 & 0.5283 \\
  CTG & 0.0000 & 0.0000 & 0.0000 & 0.0000 & 0.1250 & 0.1250 & 0.7500 \\
  GTG & 0.0000 & 0.0000 & 0.0000 & 0.0000 & 0.1322 & 0.3098 & 0.5580 \\
  TTG & 0.0000 & 0.0000 & 0.0000 & 0.0000 & 0.1333 & 0.2667 & 0.6000 \\
  TAA & 0.6414 & 0.0066 & 0.3059 & 0.0461 & 0.0000 & 0.0000 & 0.0000 \\
  TAG & 0.6030 & 0.0066 & 0.3355 & 0.0548 & 0.0000 & 0.0000 & 0.0000 \\
  TGA & 0.5991 & 0.0019 & 0.3591 & 0.0400 & 0.0000 & 0.0000 & 0.0000 \\
\end{tabular}
\end{center}
\end{description}

\end{enumerate}

\end{document}